\documentclass{article}

\usepackage{arxiv}

\usepackage[utf8]{inputenc} % allow utf-8 input
\usepackage[T1]{fontenc}    % use 8-bit T1 fonts
\usepackage[utf8]{inputenc}

\usepackage{graphicx}
\usepackage{amsmath}
\usepackage{physics}

\usepackage{hyperref}
\usepackage{multirow}

\usepackage[version=4]{mhchem}
\usepackage{longtable}

\title{A thermoelastic plate model for shot peen forming metal panels based on effective torque}

\date{} 					% Or removing it

\author{
    Conor Rowan \\
	Aerospace Engineering\\
	University of Colorado Boulder\\
	Boulder, CO 80309 \\
	\texttt{conor.rowan@colorado.edu} \\
}

\renewcommand{\headeright}{}
\renewcommand{\undertitle}{}
\renewcommand{\shorttitle}{}

\hypersetup{
pdftitle={A template for the arxiv style},
pdfsubject={q-bio.NC, q-bio.QM},
pdfauthor={David S.~Hippocampus, Elias D.~Striatum},
pdfkeywords={First keyword, Second keyword, More},
}

\begin{document}
\maketitle

\begin{abstract}
A common technique used in factories to shape metal panels is shot peen forming, where the panel is sprayed with a high-velocity stream of small steel pellets called "shot." The impacts between the hard steel shot and the softer metal of the panel cause localized plastic deformation, which is used to improve the fatigue properties of the material's surface. The residual stress distribution imparted by impacts also results in bending, which suggests that a torque is associated with it. In this paper, we model shot peen forming as the application of spatially varying torques to a Kirchhoff plate, opting to use the language of thermoelasticity in order to introduce these torque distributions. First, we derive the governing equations for the thermoelastic thin plate model and show that only a torque-type resultant of the temperature distribution shows up in the bending equation. Next, to calibrate from the shot peen operation an empirical "effective torque" parameter used in the thermoelastic model, a simple and non-invasive test is devised. This test relies only on measuring the maximum displacement of a uniformly shot peened plate as opposed to characterizing the residual stress distribution. After discussing how to handle the unconventional fully-free boundary conditions germane to shot peened plates, we introduce an approach to solving the inverse problem whereby the peening distribution required to obtain a specified plate contour can be obtained. Given that the relation between shot peen distributions and bending displacements at discrete points is non-unique, we explore a regularization of the inverse problem which gives rise to shot peen distributions that match the capabilities of equipment in the factory. In order to validate our proposed model, an experiment with quantified uncertainty is designed and carried out which investigates the agreement between the predictions of the calibrated model and real shot peen forming operations. 
\end{abstract}


%------------------------------------------------------------------------------

\section{Introduction}

\paragraph{} Shot peening is a surface treatment widely used in the aerospace industry to enhance the fatigue strength of metal structural components \cite{kirk_shot_1999}. It works by bombarding the surface of a metal part with small steel pellets called "shot" at high velocities. The impacts of the shot with the surface of the material results in localized plastic deformation and compressive residual stresses which help to prevent the initiation of cracks \cite{wu_effect_2023}. Though most aerospace components are sufficiently stiff that the residual stresses imparted by shot peening have little effect on their shape, this is not the case for thin metal panels such as wing skins \cite{badger_investigation_1981}. In fact, it is common in the aerospace industry to use shot peening specifically as a process to contour thin metal parts, as it allows for more continuous contours than a traditional procedure such as brake forming \cite{de_vin_curvature_2000}. 

\paragraph{} Historically, generating the "recipes" to form metal panels with shot peening has relied primarily on trial-and-error. The intensity of a shot peen operation describes the extent of plastic deformation in material and thus the residual stress distribution \cite{poozesh_analytical_2024, liu_characterization_2022}. As such, obtaining a shot peen forming recipe requires a map of intensity vs. position on the surface of a panel which, when carried out, obtains a desired contour of the part. In order to streamline the laborious process of obtaining these recipes through trial-and-error, a number of authors have investigated predictive models of the shot peen process. Two early approaches to modeling shot peen forming are \cite{vanluchene_numerical_1996}---where the exact characteristics of the residual stress distribution are ignored in favor of a linear elastic model incorporating stress resultants---and \cite{levers_finite_1998}, in which temperature inputs are used to model experimentally determined residual stress distributions. The benefit of the first approach is that only the net force and moment resultants of the residual stress distribution are required, but this may come at the cost of accuracy of the simulated solution. Variants of the latter approach---where a high-fidelity model of the shot peen process is used to make predictions of the deformed contour---have made up the majority of follow-up works.

\paragraph{} In the spirit of high-fidelity simulations of the shot peen process, many subsequent works have opted to directly simulate the collisions themselves in place of imposing the residual stress distribution that arises from the collisions. In \cite{wang_finite_2003}, up to 1000 impacts involving plastic deformation of the shot with the panel are directly simulated. In \cite{xiao_prediction_2018}, a smaller number of collisions are used to estimate the effects of shot peening. The authors in \cite{jebahi_robust_2016} aspire to make these impact models more scalable by coupling discrete and continuum models of the shot peen process. Another explicit collision simulation work is \cite{miao_finite_nodate}, where the shot peen process is implemented in a commercial finite element code. While highly representative of the underlying physics of the shot peen process, these collision simulations are extremely expensive due to their dynamical nature and the nonlinearity introduced by contact and plastic deformation. The computational expense also renders the inverse problem intractable. An alternative to directly simulating collisions is the so-called "eigenstrain" method, in which an experimentally characterized residual strain field can be introduced as a source term in the governing equations for stress equilibrium \cite{korsunsky_modelling_2005, faucheux_simulating_2018, miao_peen_2021}. Our review suggests that methods which use  experimental results on the through-thickness stress and/or strain distribution as source terms in a numerical simulation are the most popular. For example in \cite{levers_prediction_2014}, though not an eigenstrain method, a known residual stress distribution is introduced to the governing equations in order to predict the deformation. Some other works use thermal loads as a tool to introduce an experimentally determined plastic strain distribution \cite{wang_process_2006, zeng_finite_2003}. The experiments required to obtain the residual stress and strain profiles used in these methods require complex machinery and processing techniques \cite{gao_analysis_2002, chen_determination_2020} and thus represent a bottleneck in building predictive models of shot peen forming.

\paragraph{} Because shot peen forming recipes need to be designed on a part-by-part basis, it is important for any modeling approach to be able to solve the inverse problem, which maps a desired displacement field to a specific intensity distribution over the part's surface. In one work, the eigenstrain method is used to solve the inverse problem for the peening distribution of interest \cite{miao_shot_2022}. In \cite{sushitskii_determination_2022}, a novel non-Euclidean plate model is used to solve the inverse problem \cite{sushitskii_determination_2022}. Finally, some recent works have made use of neural networks as surrogate models for the relation between intensity and deformation, which, due to the low cost of a forward evaluation, are useful for solving inverse problems \cite{wang_process-based_2023, siguerdidjane_efficient_2020}.

\paragraph{} In this paper, we return to the ideas of one of the foundational early studies on modeling shot peen forming, where only resultants of the through-thickness stress distribution are used to predict deformation \cite{vanluchene_numerical_1996}. However, unlike past works, we rely only on the moment resultant from the through-thickness stress distribution, which can be estimated from simple measurements taken from uniformly shot peened plates. This obviates the more complex experimental efforts required to obtain the full residual stress and/or strain distribution in the peened material. Thus, our contributions in this work are as follows:

\begin{enumerate}
    \item We use thermoelasticity to derive the governing equations of a Kirchhoff plate driven by an applied torque distribution,
    \item We devise a simple, non-invasive experiment to calibrate the thermal model to the shot peen process by extracting an "effective torque,"
    \item We introduce a novel regularization of the inverse problem which encourages shot peening recipes that can be realized by existing shot peen machines,
    \item We compare the predictions of our calibrated model to experimental results obtained from the application of spatially varying shot peen distributions.
\end{enumerate}

\paragraph{} The rest of this paper is organized as follows. In Section 2, we introduce the constitutive relation for a thermoelastic material and the kinematic assumptions of Kirchhoff plate bending. With these in hand, we then derive the variational energy functional whose minimum governs the solution. In Section 3, we develop an experimental test which is used to connect the modeling framework of temperature distributions to the real shot peen operations. In Section 4, we derive the numerical solution procedure and a technique for handling the unconventional boundary conditions used in modeling shot peen forming. In Section 5, we devise a method for treating inverse problems whereby a physically-realizable peening distribution required to obtain specified displacements can be acquired. In Section 6, we show our experimental set up and the results of the experiment conducted to validate our proposed model. In Section 7, we discuss the agreement of the model predictions with the experiment and close with concluding remarks.

%------------------------------------------------------------------------------

\section{Thermoelastic plate model}

\subsection{Constitutive relation and kinematic model}

\paragraph{} By assumption, the Kirchoff plate is in a state of plane stress. The general relationship between the in-plane stress components $\sigma_{ij}$ and in-plane strain components $\epsilon_{ij}$ for an isotropic, thermoelastic material is

\begin{equation*}
\begin{aligned}
\epsilon_{11} = \frac{1}{E}\Big( \sigma_{11} -v\sigma_{22} \Big) + \alpha T,\\
\epsilon_{22} = \frac{1}{E}\Big( \sigma_{22} -v\sigma_{11} \Big) + \alpha T,\\
\gamma_{12} = 2\epsilon_{12} = \frac{1}{G}\sigma_{12},
\end{aligned}
\end{equation*}

\noindent where $E$ is the Young's modulus, $G$ is the shear modulus, $v$ is the Poisson ratio, $T$ is the applied temperature, and $\alpha$ is the coefficient of thermal expansion. The strains arise from a combination of stresses and the applied temperature. The tensorial shear strain $\epsilon_{12}$ is distinguished from its engineering counterpart $\gamma_{12}$. Inverting this relationship, stresses can be written in terms of the strain and temperature as

\begin{equation} \label{stress}
\begin{aligned}
\sigma_{11} &= \frac{E}{1-v^2}\Big( \epsilon_{11} +v\epsilon_{22} \Big)-\frac{E\alpha T}{1-v}, \\
\sigma_{22} &= \frac{E}{1-v^2}\Big( \epsilon_{22} +v\epsilon_{11} \Big)-\frac{E\alpha T}{1-v},\\
\sigma_{12} &= G\gamma_{12} = \frac{E}{2(1+v)} \gamma_{12} = \frac{E}{(1+v)} \epsilon_{12}.
\end{aligned}
\end{equation}

The shear modulus $G$ has been replaced with its definition in terms of $E$ and $v$. We employ the usual kinematic assumptions of Kirchhoff plate bending which relate strains in the plate to the derivative of the transverse displacement $u_3(x_1,x_2)$ \cite{bauchau_structural_2009}. We note that the plate is taken to lie in the $x_1-x_2$ plane, such that $x_3$ represents the out-of-plane coordinate. Using commas to denote differentiation, the in-plane displacements are $u_1(x_1,x_2,x_3) = \hat u_1(x_1,x_2) -x_3u_{3,1}$ and $ u_2(x_1,x_2,x_3) = \hat u_2(x_1,x_2)-x_3u_{3,2}$. The in-plane displacements arise both from bending and from displacement of the plate's midplane given by $\hat u_1$ and $\hat u_2$. The strain-displacement relations give rise to the following form of strain in the plate:

\begin{equation} \label{plate strains}
\begin{aligned}
\epsilon_{11} &= \frac{\partial \hat u_1}{\partial x_1}-x_3\frac{\partial^2 u_3}{\partial x_1^2}, \\
\epsilon_{22} &= \frac{\partial \hat u_2}{\partial x_2}-x_3\frac{\partial^2 u_3}{\partial x_2^2},\\
\gamma_{12} &= \frac{\partial \hat u_1}{\partial x_2}+\frac{\partial \hat u_2}{\partial x_1}  -2x_3\frac{\partial^2 u_3}{\partial x_1 \partial x_2}.
\end{aligned}
\end{equation}

Eqs. \eqref{plate strains} can be substituted into the constitutive relation given in Eqs. \eqref{stress} to write the stress components in terms of the displacement field. We now make use of these relations in forming the variational energy whose minimum governs the response of the thermoelastic plate.

\subsection{Variational energy}

\paragraph{} Given that the plate is in a state of plane stress, the stress components in the $x_3$ direction are all zero. This means that the general form of the strain energy in the plate is given by

\begin{equation*}
U = \frac{1}{2}\int_V \Big( \sigma_{11}\epsilon_{11} + 2\sigma_{12}\epsilon_{12} + \sigma_{22}\epsilon_{22}   \Big) dV.
\end{equation*}

\noindent where the integral is taken over the volume $V=[-h/2,h/2]\times[0,L_1]\times[0,L_2]$. The plate's thickness is given by $h$ and the two side lengths are $L_1$ and $L_2$. Introducing Eqs. \eqref{stress} into this expression, we obtain

\begin{equation*}
    \begin{aligned}
        U = \frac{1}{2} \int \qty( \frac{E}{1-v^2} (\epsilon_{11} + v\epsilon_{22}) - \frac{E \alpha T}{1-v} ) \epsilon_{11} +\qty( \frac{E}{1-v^2} (\epsilon_{22} + v\epsilon_{11}) - \frac{E \alpha T}{1-v} ) \epsilon_{22} + \frac{2E}{1+v}\epsilon_{12}^2 dV
    \end{aligned}
\end{equation*}

As discussed in \cite{bauchau_structural_2009}, the "in-plane" problem involving the displacement components $u_1$ and $u_2$ fully decouples from the bending problem for the transverse displacement $u_3(x_1,x_2)$. Making use of Eqs. \eqref{plate strains}, keeping only the terms involving the transverse displacement, and performing integration over $x_3$, the variational energy for the bending problem is 

\begin{equation}\label{Pi}
    \Pi =  \int_A \frac{D_1}{2}\Big(u_{3,11}^2 + u_{3,22}^2 + 2v  u_{3,11}u_{3,22}   + 2(1-v)  u_{3,12}^2 \Big) + D_2\Big( u_{3,11} + u_{3,22 }\Big) \qty( \int_{-h/2}^{h/2}  \alpha x_3 T dx_3 )dA,
\end{equation}

\noindent where the plates area is given by $A=[0,L_1]\times[0,L_2]$ and we define the constants $D_1 := \frac{Eh^3}{12(1-v^2)}$ and $D_2 := \frac{E}{1-v}$. We will also define the "thermal moment" as 

 \begin{equation}
     \tau := \int_{-h/2}^{h/2} \alpha x_3 T dx_3,
  \end{equation}

\noindent which we interpret as a measure of the effective torque associated with the through-thickness temperature distribution at each point in the plate. The response of the transverse displacement to an applied temperature distribution is found by minimizing the variational energy for bending, i.e. by computing 

\begin{equation}\label{var_energy}
    \delta \Pi = \delta \qty[\int_A \frac{D_1}{2}\Big(u_{3,11}^2 + u_{3,22}^2 + 2v  u_{3,11}u_{3,22}   + 2(1-v)  u_{3,12}^2 \Big) + D_2 \tau \Big( u_{3,11} + u_{3,22 }\Big) dA] = 0.
\end{equation}

Eq. \eqref{var_energy} is the governing equation for the thermoelastic plate bending problem we will use to model shot peen forming. The thermal moment $\tau$ is taken to be the input driving the bending response of the plate. In order for Eq. \eqref{var_energy} to model real shot peen forming operations, we need a way to connect the thermal moment to real shot peen processes. This is the focus of the following section.

%------------------------------------------------------------------------------

\section{Moment-intensity relation}

\paragraph{} We seek to design an experiment to extract the thermal moment which corresponds to shot peening a thin plate at a given intensity. This will form the link between the real shot peening operation and the thermoelastic plate model. Going forward, we will refer only to the thermal moment as opposed to the effective torque with the understanding that the former is our way of modeling the latter. We assume that the thermal moment is only a function of the intensity of the shot peening operation. This means that a uniformly peened plate will have a spatially constant thermal moment. In order to proceed, let us look at the bending moment distribution within the plate:

\begin{equation} \label{moments1}
\begin{aligned}
    M_1 = -\int_hx_3\sigma_2dx_3 = D_1\qty(\frac{\partial^2 u_3}{\partial x_2^2} + v\frac{\partial^2 u_3}{\partial x_1^2}) + D_2\tau = 0,  \\
    M_2 = \int_hx_3\sigma_1dx_3 = -D_1\qty(\frac{\partial^2 u_3}{\partial x_1^2} + v\frac{\partial^2 u_3}{\partial x_2^2}) - D_2\tau = 0, \\
    M_{12} = -\int_hx_3\sigma_{12}dx_3 =  D_1(1-v)\frac{\partial^2 u_3}{\partial x_1 \partial x_2} = 0.
\end{aligned}
\end{equation}

The bending moments associated with the through-thickness stress distribution are zero because there are no applied external forces or moments. With a constant thermal moment $\tau$ in Eqs. \eqref{moments1}, it is necessary that the transverse displacement has the following form

\begin{equation}\label{ansatz}
u_3 = Ax_1^2 + Bx_1x_2 + Cx_2^2,
\end{equation}

\noindent where we assume in this section that the origin of the coordinate system is at the center of the plate in order to avoid linear terms in the displacement of Eq. \eqref{ansatz}. This form of bending displacement is a consequence of the ansatz that a uniformly peened plate has spatially constant thermal moment. Now, plugging Eq. \eqref{ansatz} into Eqs. \eqref{moments1}, we obtain the following system of equations 

\begin{equation*}
    \begin{aligned}
    0 = D_1\Big( 2Av + 2C \Big) + D_2\tau , \\
    0 = -D_1\Big( 2A + 2Cv \Big) - D_2\tau , \\
    0 = D_1\Big(1-v\Big)B,
    \end{aligned}
\end{equation*}

\noindent which immediately shows that $B=0$. Using the definitions of $D_1$ and $D_2$, these equations can be solved to find that $A = C = -6\tau/h^3$. Thus, the relation between the displacement and thermal moment is 

\begin{equation}\label{test_disp}
    u_3(x_1,x_2) = -\frac{6}{h^{3}}\tau \Big( x_1^2 + x_2^2\Big).
\end{equation}

We now have the means of conducting an experiment to determine the connection between the shot peening operation and the thermal moment. Consider a test sample with side lengths $\ell_1$ and $\ell_2$. The thickness $h$ of the test sample should be the same as the plate we ultimately wish to model, as the thermal moment associated with shot peening may depend on the plate thickness. This is equivalent to the observation that the details of the through-thickness residual stress distribution from shot peening depends on the sample thickness \cite{gao_analysis_2002}. The geometry of the test sample is unimportant so long as it is rectangular. The maximum displacement will be obtained at the corner of the plate (given our choice of coordinate system), which allows us to calculate the relation between the thermal moment and the maximum displacement. Looking only at the magnitude of the displacement in Eq. \eqref{test_disp}, the relation between the maximum displacement, the sample geometry, and the thermal moment is

\begin{equation} \label{thermal moment}
  \tau = \frac{2h^{3}| u_{max} |}{3(\ell_1^{2}+\ell_2^{2})}  .
\end{equation} 

\paragraph{} In practice, $|u_{max}|$ is the height the sample rises to after being uniformly peened on one side. This is a quantity which is straightforward to measure experimentally, thus Eq. \eqref{thermal moment} provides a procedure to calibrate the thermoelastic model to the shot peening operation. In other words, we only need to measure the maximum displacement of a uniformly peened plate in order to extract the thermal moment corresponding to shot peening at a given intensity. Whether this line of reasoning leads to accurate predictions of deformation for non-uniform distributions of shot peen intensity remains to be seen. In general, we expect that the thermal moment depends on the plate material and thickness. Thus, in order to calculate the equivalent thermal moment associated with shot peening for a given material and thickness, a series of tests can be carried out where a plate is fully peened at different intensities and the maximum displacement is measured. This displacement is related to the moment $\tau$ and stored as in Table \ref{tab:calibration}.

\begin{table}[htbp]
\centering
\begin{tabular}{ |c|c|c| } 
 \hline
 \textbf{Intensity Level}  & $\pmb{u_{max}}$ & $\pmb{\tau}$  \\ 
 \hline
 1 & $u_1$ & $\frac{2h^{3}}{3(\ell_1^{2}+\ell_2^{2})}u_1$ \\ 
 2 & $u_2$ & $\frac{2h^{3}}{3(\ell_1^{2}+\ell_2^{2})}u_2$ \\ 
 $\vdots$ & $\vdots$ & $\vdots$ \\
 \hline

\end{tabular}
\caption{Relationship between shot peen intensity and the model input $\tau$ to be generated from tests. A table of this sort provides the link between shot peening operations and the thermoelastic model.}
\label{tab:calibration}
\end{table}

\paragraph{} The intensity $I$ of the shot peen procedure is defined as the maximum displacement of a standardized steel "almen" strip after being uniformly peened \cite{miao_numerical_2010}, so for these plates, $u_{max}=I$. Assuming there is still a linear relationship between intensity and displacement for other materials and plate thicknesses, we can write $\tau = KI$ where the constant of proportionality $K$ combines the empirical intensity-displacement and the analytical displacement-moment relationships. With this assumption of linearity, the experiments used to calibrate the model are tasked with estimating the slope $K$ for given thicknesses and materials. Note also that the sign of the thermal moment can be controlled by shot peening different sides of the plate.

%------------------------------------------------------------------------------

\section{Numerical solution procedure}

\subsection{Discretization}

\paragraph{} Having established a connection between the thermal moment $\tau$ in the governing equation of Eq. \eqref{var_energy} and shot peen forming, we now construct a numerical solution procedure to solve for the transverse displacement. We write the unknown displacement as a tensor product of 1D shape functions:

\begin{equation} \label{form} 
u_3(x_1,x_2) = \sum_{m=0}^{N-1} \sum_{n=0}^{N-1}  a_{nm}\phi^{(1)}_n(x_1) \phi^{(2)}_m(x_2).
\end{equation}

The coefficients $a_{nm}$ weight the different shape functions and there are $N$ distinct shape functions in each coordinate direction. The shape functions $\phi^{(1)}$ and $\phi^{(2)}$ each depend on one spatial coordinate and are chosen to reflect boundary conditions of the problem at hand. Eq. \eqref{form} can be written more compactly as

\begin{equation*}
    u_3(x_1,x_2) = \sum_{i=0}^{N^2-1} a_i \Phi_i = \begin{bmatrix} a_{00} \\ a_{01}  \\ \vdots \\ a_{0(N-1)} \\ a_{10} \\ a_{11} \\ \vdots \\ a_{(N-1)(N-1)} \end{bmatrix} \cdot \begin{bmatrix} \phi^{(1)}_0 \phi^{(2)}_0 \\ \phi^{(1)}_0 \phi^{(2)}_1 \\ \vdots \\ \phi^{(1)}_0 \phi^{(2)}_{N-1} \\ \phi^{(1)}_1 \phi^{(2)}_0 \\ \phi^{(1)}_1 \phi^{(2)}_1 \\ \vdots \\ \phi^{(1)}_{N-1} \phi^{(2)}_{N-1} \end{bmatrix} .
\end{equation*}

It can be seen that the entries of the new shape functions $\{\Phi_i\}_{i=1}^{N^2-1}$ comprise the two univariate shape functions in a straightforward way: 

\begin{equation} \label{phi}
\Phi_n = \phi^{(1)}_{n // N} \phi^{(2)}_{n\% N},
\end{equation}

\noindent where the notation "//" and "\%" indicates the Pythonic floor and modular division functions respectively. The other quantity in Eq. \eqref{var_energy} which we need to discretize is the thermal moment $\tau$. This can accomplished by projecting the thermal moment onto the shape functions

\begin{equation} \label{projection}
    \tau(x_1,x_2) = \sum_{n=0}^{N^2-1} t_n \Phi_n , \quad t_n = \frac{\int \tau(x_1,x_2) \Phi_n dA}{\int \Phi_n^2 dA},
\end{equation}

\noindent where we assume for simplicity that the thermal moment is discretized with the same number of shape functions as the displacement. Now, we must plug these discretized quantities into the variational energy governing the bending problem. As shown in Eq. \eqref{Pi}, the variational energy is given by

\begin{equation*} 
\Pi =  \int_A \frac{D_1}{2}\Big(u_{3,11}^2 + u_{3,22}^2 + 2v  u_{3,11}u_{3,22}   + 2(1-v)  u_{3,12}^2 \Big) +  D_2\tau \Big( u_{3,11} + u_{3,22} \Big)dA.
\end{equation*}

By substituting the discretization of the displacement and thermal moment, computing derivatives of the shape functions, and performing the required integration, we eventually obtain the discretized variational energy. See Appendix B and C for these calculations. We can write the discretized variational energy compactly as

\begin{equation} \label{quad form}
\hat \Pi = \frac{1}{2} K_{ij}a_i a_j + F_i a_i,
\end{equation}

\noindent where $\mathbf{K}$ is the stiffness matrix and $\mathbf{F}$ is the force vector. For simplicity, we take the stiffness matrix to be the symmetric part of the expression that emerges from the calculation shown in the appendices, as this leaves the computation of a minimum of the quadratic form unchanged. In other words, we have

\begin{equation*}
\begin{aligned}
    K_{ij} = \frac{1}{2}\Big( K_{ij}' + K_{ji}' \Big), \\
    K'_{ij} = D_1 \int \int \Big[ \Phi_{i,11} \Phi_{j,11}  + \Phi_{i,22} \Phi_{j,22}  + 2(1-v)\Phi_{i,12} \Phi_{j,12} + 2v\Phi_{i,11} \Phi_{j,22}  \Big] dx_1 dx_2 .
\end{aligned}
\end{equation*}

Using the projection of the applied thermal moment onto the shape functions given in Eq. \eqref{projection}, the entries of the force vector are given by 

\begin{equation} \label{force}
F_i = \sum_{j=0}^{N^2-1} D_2 t_j \int \int \Big( \Phi_{j} \Phi_{i,11} + \Phi_{j} \Phi_{i,22}  \Big) dx_1 dx_2. 
\end{equation}

The solution to the plate bending problem is governed by a minimum of Eq. \eqref{quad form} in terms of the displacement coefficients $\mathbf{a}$. This reads

\begin{equation*}
    \frac{\partial \hat \Pi}{\partial \mathbf{a}} = \mathbf{K} \mathbf{a} + \mathbf{F} = \mathbf{0}, \quad \mathbf{a} = -\mathbf{K}^{-1} \mathbf{F}.
\end{equation*}

The coefficients on the transverse displacement under specified loading conditions are straightforward to obtain once the stiffness matrix and the force vector have been constructed. These two quantities depend on the univariate shape functions $\phi^{(1)}$, $\phi^{(2)}$, and the thermal moment distribution. The shape functions are chosen to reflect the edge support configuration. We now discuss the fully-free boundary conditions relevant for shot peening and how to handle this in our numerical implementation.

\subsection{Fully-free plate}

\begin{table}[htbp]
\centering
\begin{tabular}{  |c|c| } 
 \hline
 $n$  & $P_n(x)$  \\ 
 \hline
 0 & 1 \\
 1 & $x$ \\
 2 & $(3x^2-1)/2$ \\
 3 & $( 5x^3 - 3x )/2$ \\
 4 & $(35x^4 - 30x^2 + 3)/8$ \\
 5 & $(63x^5 - 70x^3 + 15x  )/8 $ \\
 6 & $(231x^6 - 315x^4 + 105x^2 - 5)/16$ \\
 7 & $(429x^7 - 693x^5 + 315x^3 - 35x)/16$ \\
 
 \hline
 \end{tabular}
\caption{Set of 1D Legendre polynomials up to degree 7. Higher order polynomials can be taken from tabulations or calculated from closed form generating functions.}
\label{tab:legendre}
\end{table}

\begin{figure}
    \centering
    \includegraphics[scale=0.4]{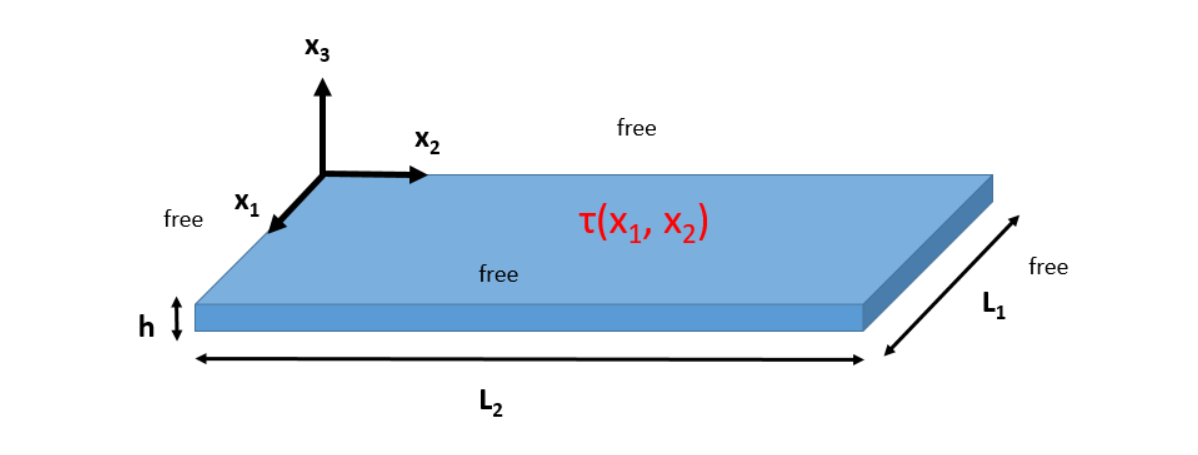}
    \caption{A fully-free plate with applied thermal moment. Given that we neglect the part's own weight, this is a model for a small peened sample resting on a table.}
    \label{FF peened plate}
\end{figure}

\paragraph{} Figure \ref{FF peened plate} shows an unconstrained plate which is subject to a thermal moment distribution. This setup models a shot peened sample resting freely on a flat surface. So long as the effect of the part's weight is small compared to the applied moments from peening, the neglect of explicit bending forces will be a realistic approximation for many panels. We choose Legendre polynomials as shape functions to handle the bending response for a plate with free edges. Thus, Eq. \eqref{phi} for the unconstrained part becomes

\begin{equation*}
    \Phi_n = P_{n//N}\qty(\frac{x_1-L_1/2}{L_1/2}) P_{n\%N}\qty(\frac{x_2-L_2/2}{L_2/2}) .
\end{equation*}

See Table \ref{tab:legendre} for the set of the first eight univariate Legendre polynomials. Because Legendre polynomials are defined on the interval $[-1,1]$, we must shift them to match our coordinate system, whose origin is at the corner of a plate with side lengths $L_1$ and $L_2$. Numerically simulating a plate with zero force and moment (fully-free) boundary conditions requires that we handle rigid body modes by manually orienting the displacement field in space, thus removing the unconstrained degrees of freedom in the displacement discretization. We will do this by zeroing the displacement at three of the four corners of the plate, i.e. by requiring that $u_3(0,0)=u_3(L_1,0)=u_3(0,L_2)=0$. These added constraints can be enforced with Lagrange multipliers. Thus, we seek a minimum of Eq. \eqref{quad form} subject to three additional constraints on the transverse displacement. Using the Legendre polynomial discretization of the displacement, the constraints read

\begin{equation*}
    \begin{aligned}
        u_3(0,0) = \sum_{n}a_n P_{n//N}(-1)  P_{n\%N}(-1) = \sum_{n}a_n (-1)^{n//N}  (-1)^{n\%N},\\
        u_3(L_1,0) = \sum_{n}a_n P_{n//N}(1)  P_{n\%N}(-1) = \sum_{n}a_n (-1)^{n\%N} ,\\
        u_3(0,L_2) = \sum_{n}a_n P_{n//N}(-1)  P_{n\%N}(1) = \sum_{n}a_n (-1)^{n//N}. 
    \end{aligned}
\end{equation*}

The properties of Legendre polynomials on the domain boundary make these expressions particularly simple. Following the Lagrange multiplier procedure, the solution is obtained by finding a stationary point of the Lagrange function given by

\begin{equation*} 
    \mathcal{L}(\mathbf{a}, \boldsymbol \lambda) = \frac{1}{2}\mathbf{a}^T \mathbf{K} \mathbf{a} + \mathbf{F} \cdot \mathbf{a} + \lambda_1 \sum_n a_n (-1)^{n//N}  (-1)^{n\%N} \\ + \lambda_2 \sum_n a_n (-1)^{n\%N} + \lambda_3 \sum_n a_n (-1)^{n//N} ,
\end{equation*}

\noindent where $\boldsymbol \lambda = [\lambda_1,\lambda_2,\lambda_3]^T$ are the Lagrange multipliers. A stationary point of the Lagrange function $\mathcal{L}$ is found by setting the coefficient and Lagrange multiplier gradients to zero. This reads

\begin{equation}\label{stationarity}
    \begin{aligned}
        \frac{\partial \mathcal{L}}{\partial a_j} = K_{ji}a_i + F_j + \lambda_1(-1)^{j//N} (-1)^{j\%N}  + \lambda_2 (-1)^{j\%N} + \lambda_3 (-1)^{j//N}  = 0,\\
        \frac{\partial \mathcal{L}}{\partial \lambda_1} = \sum_n a_n (-1)^{n//N}(-1)^{n\%N} := \mathbf{g}_1 \cdot \mathbf{a} = 0 , \\
        \frac{\partial \mathcal{L}}{\partial \lambda_2} = \sum_n a_n (-1)^{n\%N} := \mathbf{g}_2 \cdot \mathbf{a} = 0 ,\\
        \frac{\partial \mathcal{L}}{\partial \lambda_3} = \sum_n a_n (-1)^{n//N} := \mathbf{g}_3 \cdot \mathbf{a} = 0, 
    \end{aligned}
\end{equation}

\noindent where the second equality in the $\boldsymbol \lambda$ derivatives defines the vectors $\{ \mathbf{g}_i\}_{i=1}^3$. We now assemble these equations into a single system. The matrix equation for the linear system corresponding to Eq. \eqref{stationarity} is

\begin{equation} \label{system}
\begin{bmatrix} \mathbf{K} & \mathbf{G} \\ \mathbf{G}^T & \mathbf{0} \end{bmatrix} \begin{bmatrix} \mathbf{a} \\ \boldsymbol\lambda \end{bmatrix} = \begin{bmatrix} \mathbf{F} \\ \mathbf{0} \end{bmatrix} ,
\end{equation}

\noindent where the constraint matrix $\mathbf{G}$ is defined as $\mathbf{G}=[\mathbf{g}_1 , \mathbf{g}_2 , \mathbf{g}_3 ]$. Eq. \eqref{system} is the discretized governing equation for the bending response of a shot peened plate with fully-free boundary conditions discretized by a tensor product of Legendre polynomials. Because shot peened plates often rest freely on surfaces without any explicit edge supports, this is the model we will work with going forward.

%------------------------------------------------------------------------------

\section{Inverse problem}

\paragraph{} Given that shot peen forming is used as a technique to obtain pre-specified contours in metal plates, a model for the shot peen process should be well-equipped to solve inverse problems, which find a shot peen distribution that gives rise to a specified plate geometry. Until now, we have treated the thermal moment as a known input and the displacement as the unknown response. Eq. \eqref{system} is equally valid in both directions, meaning that we can input a known displacement field to solve for the required loads. However, the relationship between applied displacements and shot peen distributions is non-unique, which suggests the inverse problem needs to be regularized. We assume that the target displacement is stored at a discrete set of points. We wish to regularize the inverse problem in such a way that the peening distributions we obtain are "physical" in the sense that they can be realized on real shot peen machines. This translates to minimizing gradients of the intensity distribution, as it is difficult in practice to peen a plate with complex, spatially varying intensity distributions.

\paragraph{} We begin our treatment of the inverse problem by making the dependence of the displacement constraints on the thermal moment explicit. Given that the force vector in Eq. \eqref{force} can be decomposed into a matrix-vector product, we can rewrite Eq. \eqref{system} as

\begin{equation*} \label{eq} 
\begin{bmatrix} \mathbf{K} & \mathbf{G} \\ \mathbf{G}^T & \mathbf{0} \end{bmatrix} \begin{bmatrix} \mathbf{a} \\ \boldsymbol \lambda \end{bmatrix} = \begin{bmatrix} -\mathbf{T}^T \mathbf{t} \\ \mathbf{0} \end{bmatrix} = \boldsymbol \Gamma \mathbf{t},
\end{equation*}

\noindent where $\mathbf{T}^T$ is the matrix which takes the coefficients on the thermal moment $\mathbf{t}$ to the force vector $\mathbf{F}$ per Eq. \eqref{force}. The matrix $\boldsymbol \Gamma$ is the $-\mathbf{T}^T$ with three rows of zero appended to the bottom and it is introduced to isolate the thermal moment coefficients. Now, we introduce a truncation operator $\mathbf{S}$ which takes a vector of length $N^2+3$ to length $N^2$ by removing the final three entries. This operator has a simple matrix representation and is used to rid of the Lagrange multipliers. The displacement coefficients can then be written as an explicit function of the thermal moment:

\begin{equation} \label{a from t}
  \mathbf{a}  = \mathbf{S} \begin{bmatrix} \mathbf{K} & \mathbf{G} \\ \mathbf{G}^T & \mathbf{0} \end{bmatrix}^{-1} \boldsymbol \Gamma \mathbf{t}.
\end{equation}

Assuming that the displacement field is specified at $P$ points given by $\{ \mathbf{x}_i \}_{i=1}^P$, we use the discretization of the displacement to write the constraints as

\begin{equation*}
    \sum_{j} a_j \Phi_j(\mathbf{x}_1) = u_1 , \dots , \sum_{j} a_j \Phi_j(\mathbf{x}_P) = u_P .
\end{equation*}

This system of equations can be cast in matrix form by defining the matrix $Q_{ij} = \Phi_{j}(\mathbf{x}_i)$. The constraint equation is then $\mathbf{u} = \mathbf{Q} \mathbf{a}$ where $\mathbf{u} = [u_1,\dots,u_P]^T$. Finally, use $\boldsymbol \kappa$ to denote the inverse of the expanded system matrix in Eq. \eqref{a from t}. The constraints can then be written as a function of the moment coefficients via a series of matrix operators:

\begin{equation} \label{constraint equation}
\mathbf{u}  = \mathbf{Q} \mathbf{S} \boldsymbol \kappa \boldsymbol \Gamma \mathbf{t}.
\end{equation}

\paragraph{} Having written the displacement constraints in terms of the thermal moment, we can now introduce the regularization for the inverse problem. We seek to minimize a functional of the thermal moment $\tau$ which encourages physically realizable shot peen distributions subject to the constraints of Eq. \eqref{constraint equation}. In practice, we cannot apply arbitrarily large shot peen intensities, nor arbitrarily complex intensity distributions. The optimal shot peen distribution is both small in magnitude and as close to constant as possible. This suggests choosing a functional which penalizes large values of both $\tau$ and its derivatives. The chosen regularization functional $\mathcal{R}$ takes combinations of the square of the thermal moment and all its derivatives up to second order:

\begin{equation*} \label{R}
\mathcal{R}(\tau ) = \frac{1}{2}\int_A \Bigg[ \tau^2 + L_1L_2\qty( \qty(\frac{\partial \tau}{\partial x_1})^2 + \qty(\frac{\partial \tau}{\partial x_2})^2 ) + L_1^2 L_2^2\qty(  \qty(\frac{\partial^2 \tau}{\partial x_1^2})^2 + \qty(\frac{\partial^2 \tau}{\partial x_2^2})^2 + 2\qty(\frac{\partial^2 \tau}{\partial x_1 \partial x_2})^2  ) \Bigg] dA.
\end{equation*}

The plate side lengths are used to scale the derivatives and to normalize units. Using the discretization of the thermal moment, this expression becomes

\begin{multline} \label{Rhat}
\hat R(\tau) = \frac{1}{2} \sum_{i=0}^{N^2-1} \sum_{j=0}^{N^2-1} t_i t_j \int_A \Bigg[ \Phi_i \Phi_j + L_1L_2 \Big( \Phi_{i,1}\Phi_{j,1} + \Phi_{i,2}\Phi_{j,2}  \Big) \\ + L_1^2L_2^2\Big( \Phi_{i,11}\Phi_{j,11} + \Phi_{i,22}\Phi_{j,22} + 2\Phi_{i,12}\Phi_{j,12} \Big) \Bigg] dA,
\end{multline}

\noindent which is a quadratic form in the thermal moment coefficients. Call the matrix which specifies our measure of the thermal moment $\mathbf{H}$, so that the right hand side of Eq. \eqref{Rhat} becomes $ \mathcal{\hat R} = \frac{1}{2} \mathbf{t} \cdot \mathbf{H} \mathbf{t}$. Thus, we seek a minimum to Eq. \eqref{Rhat} subject to Eq. \eqref{constraint equation}. The method of Lagrange multipliers states that we extremize the Lagrange function given by

\begin{equation*}
\mathcal{L}(\mathbf{t}, \boldsymbol \lambda) =  \frac{1}{2}\mathbf{t}^T \mathbf{H} \mathbf{t} + \boldsymbol \lambda \cdot \Big( \mathbf{Q} \mathbf{S} \boldsymbol \kappa \boldsymbol \Gamma \mathbf{t}  - \mathbf{u} \Big),
\end{equation*}

\noindent where $\boldsymbol \lambda = [\lambda_1,\dots,\lambda_P]^T$ are the Lagrange multipliers. We set the gradient of this function with respect to the coefficients on the thermal moment and the Lagrange multipliers to zero in order to obtain the condition for a stationary point. This reads

\begin{equation*}
    \begin{aligned}
        \frac{\partial \mathcal{L}}{\partial \mathbf{t} } = \mathbf{H} \mathbf{t} + \big(\mathbf{Q} \mathbf{S} \boldsymbol \kappa \boldsymbol \Gamma \mathbf{t} \big)^T\boldsymbol\lambda = \mathbf{0}, \\
        \frac{\partial \mathcal{L}}{\partial \boldsymbol \lambda} = \mathbf{Q} \mathbf{S} \boldsymbol \kappa \boldsymbol \Gamma \mathbf{t} - \mathbf{u} = \mathbf{0}.
    \end{aligned}
\end{equation*}

The two systems of equations can be combined into a single system using block matrices with

\begin{equation} \label{inverse}
\begin{bmatrix} \mathbf{H} & \big(\mathbf{Q} \mathbf{S} \boldsymbol \kappa \boldsymbol \Gamma )^T \\ \mathbf{Q} \mathbf{S} \boldsymbol \kappa \boldsymbol \Gamma  & \mathbf{0} \end{bmatrix} \begin{bmatrix} \mathbf{t} \\ \boldsymbol \lambda \end{bmatrix} = \begin{bmatrix} \mathbf{0} \\ \mathbf{u} \end{bmatrix}.
\end{equation}

The coefficients on the thermal moment can be calculated by inverting the expanded system matrix and ignoring the values of the Lagrange multipliers. We note that the size of the discretization for the thermal moment ($N^2$) and the number of constraints on the displacement ($P$) cannot be chosen independently of each other. A solution to Eq. \eqref{inverse} may cease to exist when there are more constraints on the displacement then there are degrees of freedom in the thermal moment discretization. Eq. \eqref{inverse} defines a conceptually straightforward and computationally inexpensive approach to constructing recipes for shot peen forming of unconstrained plates.

%------------------------------------------------------------------------------

\section{Experimental setup and results}

\subsection{Overview} 

\paragraph{} We will compare predictions of our proposed thermoelastic model to real shot peen forming operations by calibrating the model on a uniformly peened plate and then testing it on spatially varying shot peen distributions. Test and measurement conditions will be chosen to reflect the conditions of an unconstrained plate. All shot peening is performed using 0.028 inch shot. A shot peen program which peens exposed surfaces at an intensity of 0.0101A will be used. As will be seen, different intensity distributions will be built up with this one program and "masking," which prevents regions from being peened by applying a rubbery tape to block the shot. Samples were cut out of a single large sheet of $1/8$ in 6061 T6 aluminum. Micrometers were used to check the thickness of the samples and 0.123 in was determined to be a more accurate estimate of the average thickness. The size of the samples used in testing was chosen to be 8 in square, which maximized the number of available samples while ensuring easily observable and measurable deformations. Samples are run through the machine on a long fixture three at a time. Double sided tape is used to attach the samples to the flat surface of the fixture. Small parts must be mounted on a fixture to avoid falling through wide grates inside the machine and firmly secured to avoid being blown away by the shot stream. The double sided tape was chosen because it is simple, it mimics free boundary conditions, and because it constrains the sample entirely from below (no clamps extending into the peened surface).

\begin{table}[htbp]
\centering
\begin{tabular}{|c|c|c|c|}
\hline
\textbf{Parameter} & \textbf{Symbol} & \textbf{Nominal Value} & \textbf{Units} \\
\hline
Side Length 1     & $L_1$           & 8.00                    & in            \\
Side Length 2     & $L_2$           & 8.00                    & in            \\
Thickness         & $h$             & 0.123                   & in            \\
Young's Modulus   & $E$             & $1 \text{e}7$         & psi           \\
Poisson's Ratio   & $v$             & 0.33                    & --            \\
\hline
\end{tabular}
\caption{Nominal values of parameters used for shot peen testing of 6061-T6 AMS 4027 aluminum sheet.}
\label{tab:shotpeen_params}
\end{table}

\subsection{Measurements}

\begin{figure}
    \centering
    \includegraphics[scale=0.29]{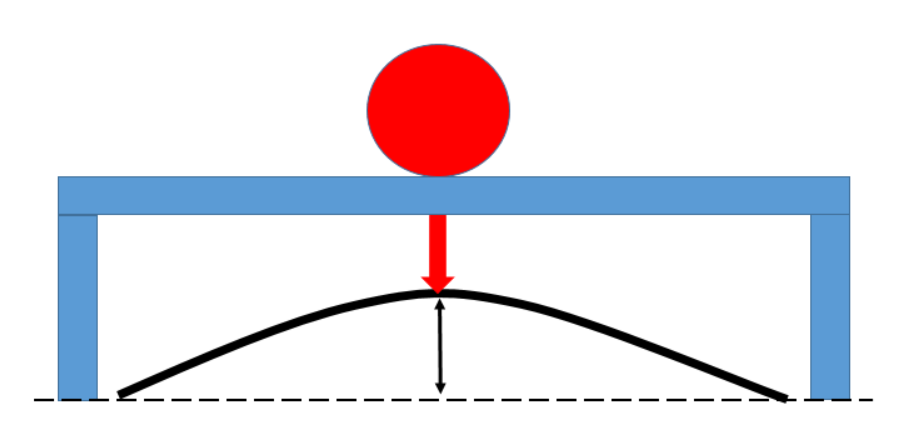}
    \caption{Schematic of height gauge measurement system. The measurement reading is the distance from the table to the top of the plate.}
    \label{measurement}
\end{figure}

\paragraph{} A number of tools are used to take the required measurements. Clamp-type micrometers are used to measure plate thicknesses and digital calipers are used for the side lengths. In order to measure the maximum displacement of the peened samples, we use a digital height gauge on a flat granite table. Figure \ref{measurement} shows a cross-section of the set-up. A spring-loaded probe extends downwards from mounting blocks of known height until contacting the formed plate. Using the dimensions of the blocks, the distance from the table to the top of the plate is measured. In practice, the approximate center of the sample will be marked and the probe will be slid around this center-point until the maximum displacement is located. This height gauge method can also be used to estimate flatness of the test plates by looking at the variation in measured height over the surface of an unpeened plate. 

\begin{figure}
    \centering
    \includegraphics[scale=0.29]{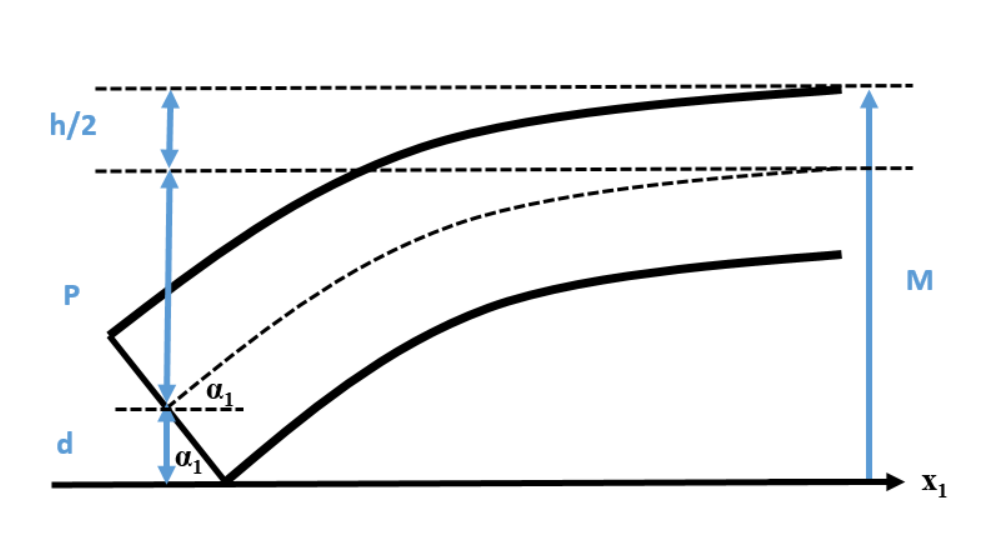}
    \caption{Cross section of a formed plate showing the relationship between the measured height $M$ and the midplane displacement $P$.}
    \label{edges}
\end{figure}

\paragraph{} We note that the plate bending model deals with the displacement of the midplane of the plate, thus the height gauge does not measure the same quantity that the numerical solution predicts. As we will see, a simple relationship governs the relationship between the predicted and measured quantities. Figure \ref{edges} shows an $x_2$ slice of a plate with thickness $h$ with the midplane shown as a dashed line. By the Kirchoff plate assumptions, the normal material line remains perpendicular to the midplane as shown at the edge. The measurement $M$ goes from the table to the top of the plate, whereas the model prediction $P$ is the displacement of the midplane. The distance $d$ from the table to the midplane at the plate's edge is less than half the thickness because of the rotation of the edge. As Figure \ref{edges} shows, the model prediction and the measurement are related by

\begin{equation*}
    d + P + h/2 = M .
\end{equation*} 

Now, note that the edge will be rotated around both the $x_1$ and $x_2$ axes, so that there exists a second angle of rotation which is not shown in the figure. In practice, these rotation angles will be very small. This allows us to say that $d\approx h/2$. The relationship between the model's prediction of the midplane displacement and the measurement is then

\begin{equation} \label{relation}
P = M - h.
\end{equation}

\subsection{Experimental design}

\begin{figure}
    \centering
    \includegraphics[scale=0.45]{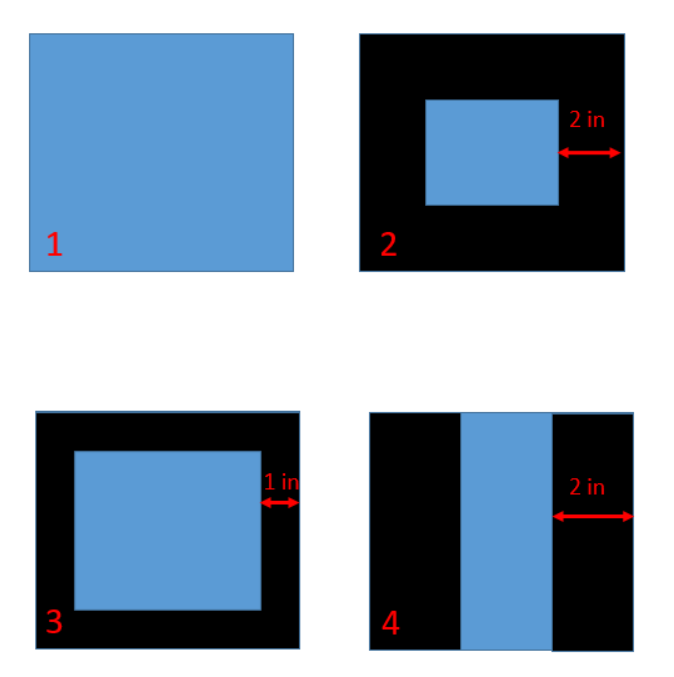}
    \caption{The shot peen distributions used in the tests are numbered for reference. Black indicates a masked region of the sample whereas blue is peened. Masking is accomplished with one and two inch tape. Configuration 1 is used to calibrate the thermal model whereas the other configurations are used as comparisons to predictions of the calibrated thermoelastic model.}
    \label{configs}
\end{figure}

\paragraph{} Before we can use the thermal model to make predictions, samples must be uniformly peened to extract the thermal moment corresponding to shot peening per Eq. \eqref{thermal moment}. With shot peening at 0.0101A intensity and masking, we construct three non-uniform intensity distributions by which the model can be assessed. Parsimony governed the choice to use a single shot peen program and intensity---both from the standpoint of testing logistics and the forthcoming uncertainty quantification. The configurations used in the tests are shown in Figure \ref{configs}. The rubber tape used for masking is available in one and two inch widths, so these configurations were selected to be convenient and repeatable. Note that neither the masking nor the tape used to hold the samples on the fixture introduces constraints on the displacement of the samples used in the experiment. In particular, the double sided tape is not so strong as to prevent the part from rising off the fixture at its center when peened.

\paragraph{} Having decided on the nature of the samples, the method of peening, the holding fixture, and the intensity distributions to be tested, we turn to the structure of the experiment itself. Because the fixture will have to be run multiple times to accommodate a realistic number of samples, run-to-variation in the shot peen conditions is introduced. Furthermore, prior knowledge of the machine suggests that the position on the holding fixture potentially influences the shot peen intensity a sample sees within a given run. Our task is to design an experiment which is informed by the hypothesized sources of variation. Because the predictions of our model are driven entirely by estimates of the thermal moment, we seek confidence in our estimate of this parameter---both in its value and associated uncertainty. A run with all three positions occupied by uniformly peened plates will yield an estimate of position-induced variation and three runs at a given position will yield an estimate of run-to-run variation. Thus, we choose to perform three runs of uniformly peened plates at each of the three positions to better understand the underlying variation in the experiment. These runs of uniform peening will be at the beginning, middle, and end of the experiment. We choose to carry out the experiment in a single session to further reduce the confounding effects of potential machine-induced variation. Interspersed with the uniform plates are the masked configurations, each of which will be replicated three times. This totals to 6 runs through the machine and 18 samples. Though we are able to estimate run-to-run and position variation separately for the masked samples, these two sources of uncertainty will be effectively lumped together by running the masked samples under maximally variable conditions. In other words, no two replicates of a certain configuration are peened in the same run or at the same position between runs. See Table \ref{tab:configs} for a summary of the experimental design.

\begin{table}[htbp]
\centering
\begin{tabular}{  |c|c|c|c| } 
\hline
\textbf{Sample} & \textbf{Run}  & \textbf{Position} & \textbf{Configuration}    \\ 
\hline
1 & 1 & 1 & 1 \\
2 & 1 & 2 & 1 \\
3 & 1 & 3 & 1 \\
\hline
4 & 2 & 1 & 4 \\
5 & 2 & 2 & 3 \\
6 & 2 & 3 & 2 \\
\hline
7 & 3 & 1 & 2 \\
8 & 3 & 2 & 4 \\
9 & 3 & 3 & 3 \\
\hline
10 & 4 & 1 & 1 \\
11 & 4 & 2 & 1 \\
12 & 4 & 3 & 1 \\
\hline
13 & 5 & 1 & 3 \\
14 & 5 & 2 & 2 \\
15 & 5 & 3 & 4 \\
\hline
16 & 6 & 1 & 1 \\
17 & 6 & 2 & 1 \\
18 & 6 & 3 & 1 \\
\hline
\end{tabular}
\caption{Each sample is given a number for identification. The experiment is split into blocks by run number.}
\label{tab:configs}
\end{table}

\subsection{Uncertainty quantification}

\begin{figure}
    \centering
    \includegraphics[scale=0.5]{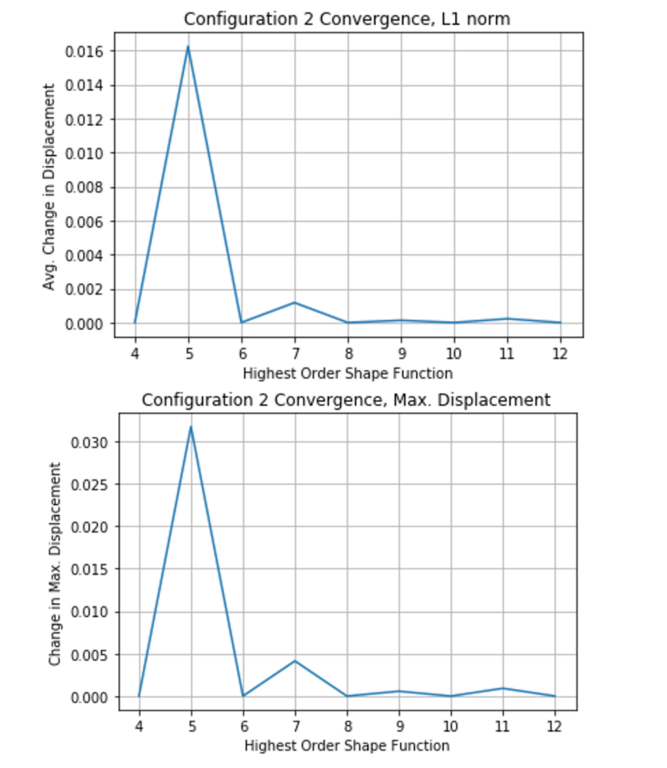}
    \caption{Convergence of the model to itself. The plots show two measures of change in the displacement field as the highest order of the two shape functions increases. After degree 8, refining the approximation has little effect on the solution by both measures. Nearly identical behavior is observed for Configurations 3 and 4.}
    \label{convergence}
\end{figure}

\paragraph{} To better compare theoretical predictions to the test samples, we seek to propagate uncertainty in model inputs through the solution procedure to obtain uncertainty bounds on the model predictions. This will estimate the degree to which realistic deviations of problem parameters from their nominal values influence the displacement predictions. To do this, we measure/estimate the variation in the uncertain inputs and devise a method to account for this. We begin by listing and discussing the various sources of uncertainty:

\begin{itemize}
    
    \item \textbf{Side lengths: }samples are cut with a shear machine out of sheet stock. Measuring dimensions with digital calipers reveals the extent of the variation between samples. It is assumed that the two side lengths are independent of one another. The effect side length variation can be accounted for by running the model with different values of $L_1$ and $L_2$.
    
    \item \textbf{Thickness: }the thickness can be measured within and among samples to estimate variability. Clamping micrometers yield the desired accuracy and provide measurements which do not confound thickness and flatness. The thickness of the plate, which is assumed to only have variation among samples (no spatial variation), can be accounted for by inputting it directly into the thermal model.
    
    \item \textbf{Thermal moment: }this parameter is the bridge between shot peening and the thermoelastic model. Its value is calculated with the uniformly peened test plates and Eq. \eqref{thermal moment}. The measured displacements of the uniformly peened plates in the experiment will be influenced by run-to-run and positional variation in the machine. They may also be influenced by variability in the material properties of the plate, the condition of its surface, and other unaccounted for sources of variation. Many variables which cannot be independently measured will be lumped into the distribution of the measured displacements of the uniformly peened plates. The thermal moment calculation also makes use of the side lengths and thickness, which are uncertain to the extent that we have not kept track of the dimensions of particular samples. As we will discuss further, the variation in the thermal moment will be estimated with Eq. \eqref{thermal moment} by sampling the thickness, side lengths, and max displacement independently. 
    
    \item \textbf{Masking: }the size of the masked area can vary as a result of inconsistencies in applying the tape. The distance from the edge of the sample to the unmasked area can be measured with digital calipers on a variety of samples. This phenomenon is driven by the tape being imperfectly indexed to the sample's edge and by stretching effects. The size of the masked region influences the thermal moment distribution which drives displacement in the model. The width of the masked region can be accounted for as a model input.
    
    \item \textbf{Measurement: }a study can be conducted using the height gauge setup of Figure \eqref{measurement} where the same measurements of peened samples are made by multiple individuals and recorded. The discrepancy in these measurements will be considered measurement error. Measurement errors are not model inputs but act to introduce uncertainty around a model prediction for a given set of inputs. This can be accounted for by introducing random noise to the model's prediction.
    
    \item \textbf{Flatness: }the height gauge can also be used to measure the flatness of unpeened samples. Probing around the plate, we look for heights which are larger than the plate's thickness. It was observed that some samples had significant flatness deviations at their corners ($\approx 0.010$ in), likely from being cut on the shear machine. Unpeened plates are assumed to be flat and stress-free in the thermal model, so flatness is not an input. We do not have a way to account for the effect of flatness deviations on the model's predictions.
    
    \item \textbf{Model convergence: }a finite series is used to represent the thermal moment distribution and the displacement field, thus truncation error is introduced. The extent of convergence error can be estimated by observing the change in the displacement field under the shot peen conditions of Configuration 2 as the number of shape functions increases. The highest order 1D Legendre polynomial available to use in our discretization is degree 12, so negligibly small changes in the displacement field before degree 12 indicate adequate convergence. See Figure \ref{convergence} for a convergence study. For the given applied thermal moment, we take the model to be converged after order 8 of the 1D Legendre polynomial used in the tensor product discretization of the displacement. In general, relatively low frequency shape functions accurately represent the displacement because the fourth order Kirchhoff bending model has a significant smoothing effect on the inputs.
    
    \item \textbf{Assumptions: }we have assumed that the 6061 T6 aluminum is isotropic, that the material constants are known exactly, that the tensile and compressive modulus are equivalent, that the thickness is constant, that small flatness deviations do not meaningfully affect the physics of bending or our measurements, that shot peening does not change the thickness of the sample, that changes in sample length dimensions from peening are negligible, that overspray shot peen on the edges of the plate has no effect, and most importantly, that plasticity can be neglected by modeling shot peen forming as the application of distributed torques. As we either have not or cannot account for the effect of these assumptions, the comparison of the experimental results to model predictions will determine to what degree they are reasonable. 
    
\end{itemize}

\paragraph{} We will numerically generate a distribution of predictions by using Monte Carlo methods. Namely, the model is used to repeatedly predict the displacement at the center of the sample with the inputs whose variation we can account for sampled from probability distributions. This process is repeated a set number of times and a distribution of predictions is made for each intensity configuration used in the tests. See Table \ref{tab:errors} for the sources of variation which are accounted for in this way. It is assumed that each parameter is uniformly distributed around its nominal value with a width given by the errors in the table. Note that the side length and thickness values used to calculate the thermal moment will be sampled independent of the dimensions of the plate whose displacement we  predict in each Monte Carlo trial. The displacement used in Eq. \eqref{thermal moment} will be sampled from a uniform distribution encompassing the full range of displacements measured off the uniformly peened plates. Thus, unlike the other uncertain parameters, we do not have a closed form expression for the distribution of the thermal moment.

\begin{table}[htbp]
\centering
\begin{tabular}{  |c|c|c| } 
\hline
\textbf{Parameter} & \textbf{Nominal}  & \textbf{Error}    \\ 
\hline
Thermal Moment ($\tau$) & -- & -- \\
Side Length 1 ($L_1$)& 8.000 & $\pm$ 0.025  \\
Side Length 2 ($L_2$)& 8.000 & $\pm$ 0.025  \\
Thickness ($h$)& 0.123 & $\pm$ 0.0015 \\
Masking & 0.000 &$\pm$ 0.050  \\
Measurement & 0.000 & $\pm$ 0.001 \\

\hline
\end{tabular}
\caption{Sources of variation accounted for in the Monte Carlo method of quantifying uncertainty in the model predictions. The nominal value and error of the thermal moment depend on the results of the tests and Eq. \eqref{thermal moment}. Masking error refers to the deviation from the nominal tape widths used to generate the spatially varying intensity distributions of Configurations 2, 3, and 4. }
\label{tab:errors}
\end{table}

\begin{figure}
    \centering
    \includegraphics[scale=0.12]{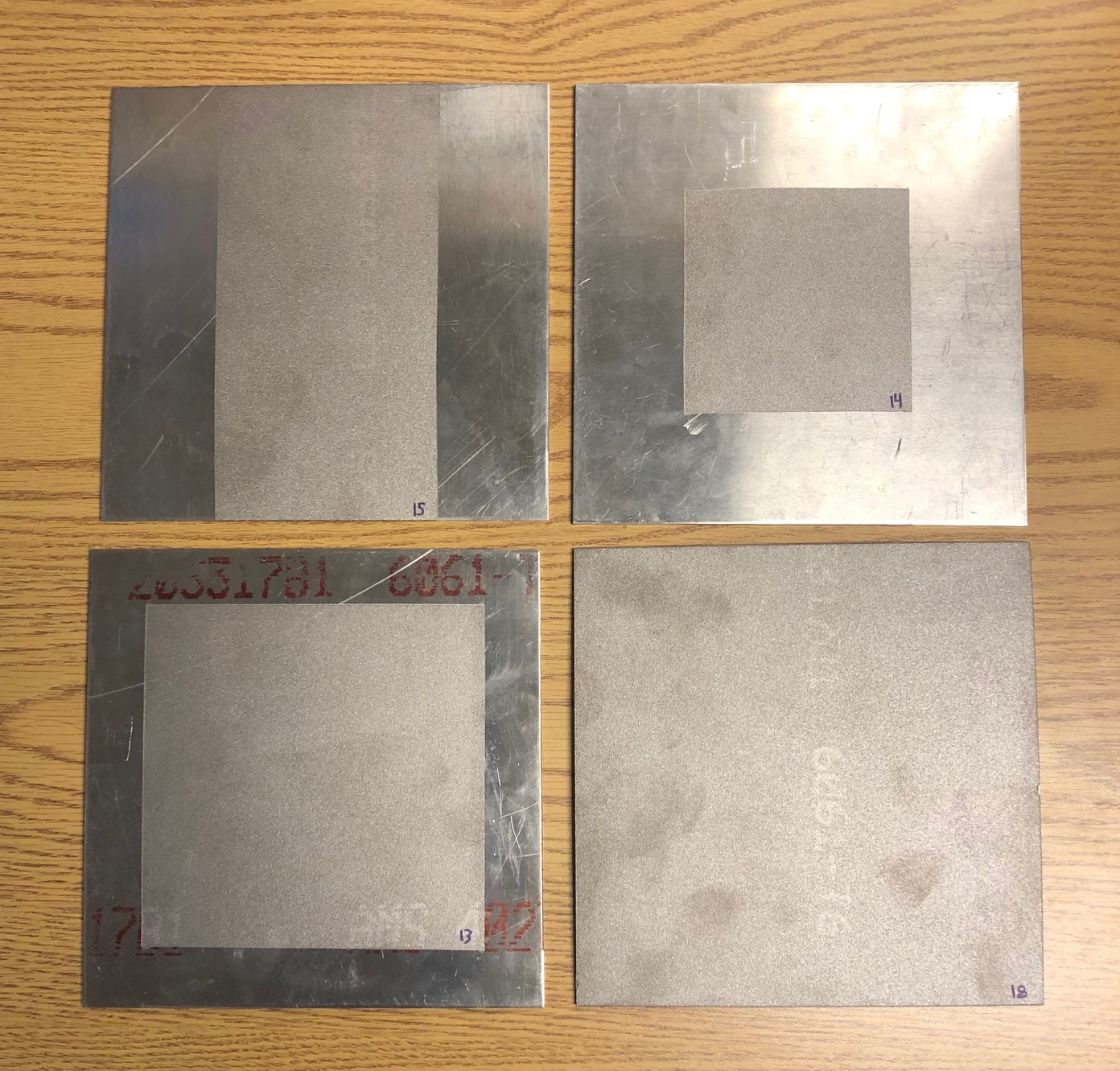}
    \caption{Appearance of the samples when peened per the four configurations used in the experiment. }
    \label{samples}
\end{figure}

%------------------------------------------------------------------------------

\subsection{Results}

\begin{table}[htbp]
\centering
\begin{tabular}{  |cc|ccc| } 
\hline
& & \multicolumn{3}{c|}{Position} \\
 & & 1 & 2 & 3  \\ 
\hline
\multirow{3}{2em}{Run} & 1 & 0.311 & 0.302 & 0.311 \\
& 4 & 0.301 & 0.300 & 0.300  \\
& 6 & 0.304 & 0.306 & 0.307 \\
\hline
\end{tabular}
\caption{Measurements taken from the height gauge of the nine uniformly peened plates used to calculate the thermal moment and estimate machine variability. This data indicates that positional variation is less than run-to-run variation. }
\label{tab:measurements}
\end{table}

\paragraph{} An discussion of the test results begins with the uniformly peened plates. Table \ref{tab:measurements} summarizes the measurements taken from these nine samples. Note that the measurements are not midplane displacements, rather the quantity $M$ in Eq. \eqref{relation} taken directly from the height gauge. Going forward, we will call this the "measured height." A two-way ANOVA with no replication can be performed for the effects of the run \# (rows) and position (columns). At a 10\% significance level, the null hypothesis of no row effect is rejected, indicating the existence of the assumed run-to-run variation. The column effect is not significant, meaning this data suggests the position on the fixture has no meaningful effect. The maximum displacement used to calculate the thermal moment with Eq. \eqref{thermal moment} will be generated by sampling a thickness value $h \sim \mathcal{U}(0.1215,0.1245)$, and subtracting it from a measurement sampled from $M \sim \mathcal{U}(0.302,0.311)$. This same thickness, in addition to the independently sampled length dimensions $L_1,L_2 \sim \mathcal{U}(7.975,8.025)$, will be used to calculate the thermal moment per Eq. \eqref{thermal moment} at the beginning of each Monte Carlo trial. Similarly, the dimensions of the sample and the size of the masked region will be sampled to predict the midplane displacement through the numerical solution. The thickness used in this calculation will be added to the midplane displacement per Eq. \eqref{relation} along with a measurement error. By repeating this process, we generate an empirical distribution of predicted measured height values which can be compared with the test results. From numerical experimentation, we observed that 250 samples ensures that the mean and standard deviation of the Monte Carlo distribution stabilize within 2\%. This is true for each of the three non-uniform shot peen configurations. 

\begin{figure}
    \centering
    \includegraphics[scale=0.58]{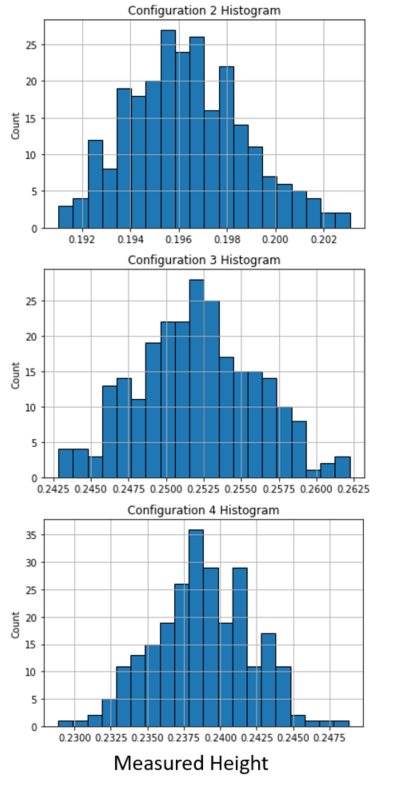}
    \caption{Monte Carlo distributions of the three peening configurations generated using 250 samples.}
    \label{histograms}
\end{figure}

\begin{figure}
    \centering
    \includegraphics[scale=0.55]{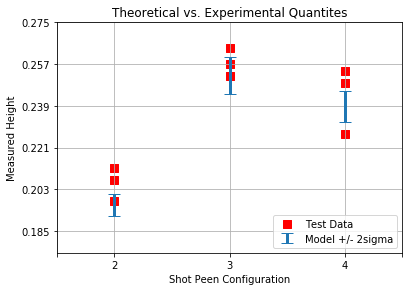}
    \caption{Overlay of the model predictions with quantified uncertainty and results from the experiment.}
    \label{results}
\end{figure}

\paragraph{} With this method of accounting for the uncertainty of the model predictions, we are ready to assess the model against the measurements taken from the non-uniform samples. These measurements are shown in Table \ref{tab:results}. See Figure \ref{histograms} for the distributions of the thermoelastic model's prediction of the measured heights and Figure \ref{results} for the comparison of these distributions to the results of the experiment. The mean and standard deviation are taken from each of the Monte Carlo distributions and the error bars in Figure \ref{results} are $\pm$ two standard deviations from the mean.

\begin{table}[htbp]
\centering
\begin{tabular}{  |c|c|c| } 
\hline
\textbf{Config.} & \textbf{Sample \#} & \textbf{Measurement}  \\
\hline
\multirow{3}{2em}{2} & 6 & 0.207 \\
& 7 & 0.212 \\
& 14 & 0.198 \\
\hline
\multirow{3}{2em}{3} & 5 & 0.252\\
& 9 & 0.264 \\
& 13 & 0.257 \\
\hline
\multirow{3}{2em}{4} & 4 & 0.254\\
& 8 & 0.249 \\
& 15 & 0.227 \\
\hline
\end{tabular}
\caption{Measurements taken from height gauge of the samples with non-uniform intensity distributions. These values are used in assessing the validity of the proposed thermoelastic model.}
\label{tab:results}
\end{table}

\subsection{Inverse problem results}

\begin{table}[htbp]
\centering
\begin{tabular}{  |c|c|c| } 
\hline
\multicolumn{3}{|c|}{Specified displacements} \\
\hline
$\pmb{x_1}$ & $\pmb{x_2}$  & $\pmb{u_3}$ \\ 
\hline
8.00 & 8.00 & 0.000 \\
4.00 & 4.00 & 0.050 \\
\hline

\end{tabular}
\caption{A simple test of the inverse method for the unconstrained plate. In this case, we zero the fourth corner of the plate and specify a single displacement at the center. We test whether the chosen regularization finds an intensity distribution that is approximately constant.}
\label{tab:invserse_problem_tab}
\end{table}

\begin{figure}
    \centering
    \includegraphics[scale=0.40]{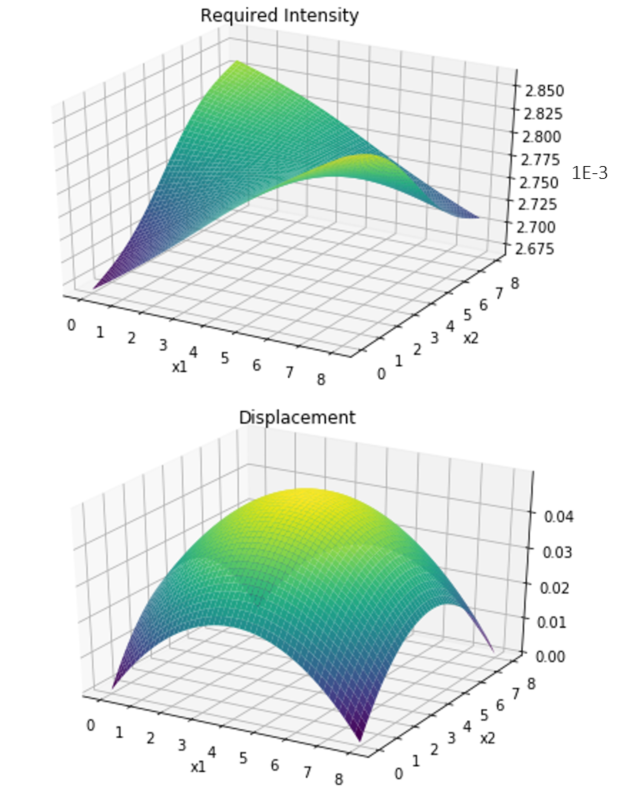}
    \caption{The intensity distribution calculated with the regularization of Eq. \eqref{Rhat} is smooth and approximately constant. Obtaining distributions of this sort is important if they are to be implemented by shot peen machines.}
    \label{inverse_fig}
\end{figure}

\paragraph{} We claim that our method for the inverse problem does not introduce new physics to the thermoelastic plate model and we thus restrict ourselves to a brief discussion of its implementation along with a numerical example. In other words, we can rely on the results of the previous section in assessing the likelihood of generating the specified displacements from the predicted thermal moments. Of primary importance for the inverse problem is the choice of regularization, which should favor shot peen distributions which can be implemented by real shot peen machines. Most shot peen machines will have limited ability to realize intensity distributions with large spatial gradients, thus the regularization of the inverse problem should encourage such slow-varying distributions. This is our rationale for choosing the regularization given in Eq. \eqref{Rhat}. Table \ref{tab:invserse_problem_tab} shows the most straightforward test of the inverse method---we seek a recipe for peening a sample which gives rise to a certain arc height when resting on a flat surface. Three of the four corners are already zeroed from the Lagrange multipliers, but we must artificially impose the zeroing of the fourth corner. Considering the test results in the previous section, 0.050 in midplane displacement at the center is small and should require a lower intensity than was used in the experiments. However, we do not know the general moment-intensity relation, as it was only calculated at the experimental intensity of 0.0101A. It is clear that zero intensity gives rise to zero displacement, so we will assume a linear relationship for the sake of this example. This allows us to write

\begin{equation*}
    \tau = \frac{2h^3}{3(L_1^2+L_2^2)}u_{max}(I)=\frac{2h^3}{3(L_1^2+L_2^2)} \frac{0.182}{0.0101}I, 
\end{equation*}

\noindent where $0.182$ is a representative midplane displacement from the test results shown in Table \ref{tab:measurements} and the dimensions of the sample are treated as exact. Figure \eqref{inverse_fig} shows the results of the inverse method using the inputs specified in Table \ref{tab:invserse_problem_tab} and the regularization functional given in Eq. \eqref{Rhat}. The shape of the displacement field is approximately spherical and satisfies the imposed constraints. The intensity distribution is smooth and minimally variable. These results suggest that uniformly peening the sample at 0.003A intensity would generate a midplane displacement $\approx$ 0.050 in.

%------------------------------------------------------------------------------

\section{Conclusion}

\paragraph{} We have hypothesized a model for shot peen forming which allows plasticity to be neglected and that avoids reliance on complex experiments to obtain the residual stress/strain distribution in the material. In particular, we hypothesized that shot peen forming can be modeled with spatially varying applied torques. We opted to use the modeling framework of thermoelasticity in order to introduce these spatially varying torques. With the caveat that this relationship is dependent on both the material and thickness of the plate, a test was devised to extract the effective torque corresponding to shot peening at a given intensity. Then, a numerical procedure was developed and implemented in order to solve the thermoelastic equations for the fully-free boundary conditions relevant to shot peen forming. An inverse method was formulated in order to generate shot peen forming recipes from the desired contour of a plate. Finally, an experiment was designed by which to test the proposed thermoelastic model and uncertainty quantification was used to carefully compare model predictions to the results of the experiment. Figure \ref{results} shows the comparison of our model to measurements taken from the test samples. We see that three values of the measured height of the test samples fall within the two standard deviation range of uncertainty on the model predictions. Five measurements were larger than the model predictions and one smaller. The data suggests that the model tends to underestimate the experimental measurements, though for Configuration 4 especially, we observe significant variation underlying the shot peen process. Taking the average of the test data and comparing to the mean theoretical prediction, the model exhibits 10\% relative error. More testing will be required to understand if the observed patterns are true systematic shifts in the model or simply artifacts of variation in the shot peen process. This will involve using different sample dimensions, materials, and further replication. Further experimentation might also focus on validating/implementing the inverse method. Finally, in order to extend this model to larger plates, it is necessary to incorporate a distributed bending force to the model to account for the part's own weight. In order to extend the effective torque approach to existing thermoelastic finite element codes, a through-thickness temperature distribution can be obtained from the measured thermal moment, which is discussed in Appendix D. Such an extension to finite element analysis would enable the modeling of peened plates with more complex geometries. 

%------------------------------------------------------------------------------

\section{Acknowledgments}

\paragraph{} This paper is a revised version of work conducted at the Frederickson Skin \& Spar facility of Boeing Commercial Airplanes (BCA) in 2021-2022. This paper contains no proprietary or confidential information and was not officially sponsored by BCA. I would like to thank the operators of the shot peen machine who helped me to obtain experimental results.

%------------------------------------------------------------------------------

\begin{appendix}

\section*{Appendix A --- Orthogonality relations} 

\paragraph{} This Appendix shows the orthogonality relations for the shifted Legendre polynomials which we use in the derivation of the governing equations:

\begin{equation*}
\begin{aligned}
    \int_0^L P_n \qty(\frac{x-L/2}{L/2}) P_m \qty(\frac{x-L/2}{L/2}) dx = \frac{L}{2} \int_{-1}^1 P_n(u) P_m(u) du \\
    = \frac{L}{2n+1}\delta_{nm} := \delta_P(n,m).
\end{aligned}
\end{equation*}

\section*{Appendix B --- Term calculations}

\paragraph{} This Appendix shows calculations of the terms in the discretized variational energy for thermoelastic plate bending:

\begin{equation*}
    \int u_{3,11}^2 dA = \sum_{i=0}^{N^2-1} \sum_{j=0}^{N^2-1} a_i a_j \int \int \Phi_{i,11} \Phi_{j,11} dx_1 dx_2.
\end{equation*}

\begin{equation*}
    \int u_{3,22}^2 dA = \sum_{i=0}^{N^2-1} \sum_{j=0}^{N^2-1} a_i a_j \int \int \Phi_{i,22} \Phi_{j,22} dx_1 dx_2 .
\end{equation*} 

\begin{equation*}
    \int u_{3,12}^2 dA = \sum_{i=0}^{N^2-1} \sum_{j=0}^{N^2-1} a_i a_j \int \int \Phi_{i,12} \Phi_{j,12} dx_1 dx_2.
\end{equation*} 

\begin{equation*}
    \int u_{3,11} u_{3,22} dA = \sum_{i=0}^{N^2-1} \sum_{j=0}^{N^2-1} a_i a_j \int \int \Phi_{i,11} \Phi_{j,22} dx_1 dx_2.
\end{equation*}

\begin{equation*}
    \int \tau \Big( u_{3,11} + u_{3,22} \Big) dA = \\ \sum_{i=0}^{N^2-1} \sum_{j=0}^{N^2-1} t_i a_j \int \int \Big( \Phi_{i} \Phi_{j,11} + \Phi_{i} \Phi_{j,22} \Big) dx_1 dx_2. 
\end{equation*}

\section*{Appendix C --- Integral calculations}

\paragraph{} This appendix shows how orthogonality can be used to expedite populating the stiffness matrix and force vector. Orthogonality of the shape functions is used so that zero entries can be recognized without carrying out the integrals. Integrals are computed over $(x_1,x_2) \in [0,L_1] \times [0,L_2]$ and the integrals factorize because the shape functions have a tensor product structure. Note that derivatives of Legendre polynomials are not themselves Legendre polynomials, thus orthogonality does not apply to them. The following manual calculations are used to expedite the numerical implementation:

\begin{equation*}
    \begin{aligned}
    \int \Phi_{i,11} \Phi_{j,11} dA = \int \frac{\partial^2}{\partial x_1^2} P_{i//N} \frac{\partial^2}{\partial x_1^2} P_{j//N} dx_1 \int P_{i\%N} P_{j\%N} dx_2 \\
    = \qty(\int \frac{\partial^2}{\partial x_1^2} P_{i//N} \frac{\partial^2}{\partial x_1^2} P_{j//N} dx_1) \delta_{P}(i\%N, j\%N) .
    \end{aligned}
\end{equation*}

\begin{equation*}
    \begin{aligned}
         \int \Phi_{i,22} \Phi_{j,22} dA = \int P_{i//N} P_{j//N} dx_1 \int \frac{\partial^2}{\partial x_1^2} P_{i\%N} \frac{\partial^2}{\partial x_1^2} P_{j\%N} dx_2 \\
         = \delta_{P}(i//N, j//N) \qty(\int \frac{\partial^2}{\partial x_1^2} P_{n\%N} \frac{\partial^2}{\partial x_1^2} P_{i\%N} dx_2).
    \end{aligned}
\end{equation*}

\begin{equation*}
     \int \Phi_{i,11} \Phi_{j,22} dA = \int \frac{\partial^2}{\partial x_1^2}P_{i//N} P_{j//N} dx_1 \int P_{i\%N} \frac{\partial^2}{\partial x_2^2} P_{j\%N} dx_2.
\end{equation*}

\begin{equation*}
    \begin{aligned}
        \int \Phi_i \Phi_j dA = \int P_{i//N} P_{j//N} dx_1 \int P_{i\%N} P_{j\%N} dx_2 \\
         = \delta_P(i//N,j//N) \delta_P(i\%N,j\%N).
    \end{aligned}
\end{equation*}

\begin{equation*}
    \begin{aligned}
        \int \Phi_i \Phi_{j,11} dA = \int P_{i//N} \frac{\partial^2}{\partial x_1^2}P_{j//N} dx_1 \int P_{i\%N} P_{j\%N} dx_2  \\
        =\qty( \int P_{i//N} \frac{\partial^2}{\partial x_1^2}P_{j//N} dx_1) \delta_P(i\%N,j\%N).
    \end{aligned}
\end{equation*}

\begin{equation*}
    \begin{aligned}
        \int \Phi_i \Phi_{j,22} dA = \int P_{i//N} P_{j//N} dx_1 \int P_{i\%N} \frac{\partial^2}{\partial x_2^2} P_{j\%N} dx_2 \\
        = \delta_P(i//N,j//N)  \qty(\int P_{i\%N} \frac{\partial^2}{\partial x_2^2} P_{j\%N} dx_2).
    \end{aligned}
\end{equation*}

\section*{Appendix D --- Thermal finite element analysis}

\paragraph{} Working with rectangular plates, it is possible to use spectral basis functions to obtain a solution. On more complicated plate geometries, it will be necessary to employ the finite element method. This raises the question of how the method we have developed can be transferred to existing thermoelastic finite element codes. To do this, we can calculate a temperature distribution that gives rise to the thermal moment associated with shot peening at a given intensity. By definition,

\begin{equation*}
    \tau = KI = \int_{-h/2}^{h/2}x_3\alpha T dx_3.
\end{equation*}

We want to find the temperature distribution $T(x_3)$ which gives rise to the known thermal moment $\tau$. The moment-intensity tests/results are used to associate the thermal moment and shot peen intensity via the constant $K$. Obviously, there isn't a unique solution as the temperature could take many forms while leaving this integral unchanged. For simplicity, we can assume that the temperature distribution is linear about the midpoint of the plate, i.e. $T(x_3)=T_0x_3$. We can then calculate the slope $T_0$:

\begin{equation*}
    \tau = \int_{-h/2}^{h/2}\alpha T_0 x_3^2 dx_3 = \frac{\alpha T_0h^3}{12} \implies T_0 = \frac{12KI}{\alpha h^3}.   
\end{equation*}

As expected, the slope of the temperature distribution increases with the shot peen intensity. Though the choice of a linear temperature distribution is arbitrary, it might be justified as a way to enforce the assumption of strain which is linearly distributed through the plate thickness as the Kirchhoff model requires. Thus, an existing thermoelastic finite element code can predict displacements from a given map of shot peen intensity by applying the appropriate temperature distribution at each point in the plate. 

\end{appendix}

%------------------------------------------------------------------------------

\bibliographystyle{plain}
\bibliography{My_Library}

\end{document}